\documentclass[12pt]{iopart}

\pdfoutput=1

\usepackage[english]{babel}
\usepackage{
amssymb,amsfonts,latexsym}
\usepackage{graphicx}
\usepackage{color}
\usepackage{iopams}
\usepackage{subfigure}
\usepackage{afterpage}
\usepackage{blkarray}
\usepackage{cite}
\usepackage{tikz}
\usetikzlibrary{snakes}
\usetikzlibrary{calc}
\usepackage{bm}
\usepackage{braket}

\definecolor{blue(pigment)}{rgb}{0.2, 0.2, 0.6}
\usepackage{mathrsfs}
\usepackage{times}
\usepackage[colorlinks,
citecolor=blue,linkcolor=blue,urlcolor=blue]{hyperref}
\usepackage{scalerel}
\usepackage{tensor}
\usepackage{etoolbox}

\makeatletter
\def\@mkboth#1#2{}
\newlength\appendixwidth
\preto\appendix{\addtocontents{toc}{\protect\patchl@section}}
\newcommand{\patchl@section}{%
  \settowidth{\appendixwidth}{\textbf{Appendix }}%
  \addtolength{\appendixwidth}{1.5em}%
  \patchcmd{\l@section}{1.5em}{\appendixwidth}{}{\ddt}%
}
\makeatother

\def\eqref#1{(\ref{#1})}
\usepackage{cancel}

\newtheorem{prop}{Proposition}

\newcommand{\be}{\begin{equation}}
\newcommand{\ee}{\end{equation}}
\newcommand{\bea}{\begin{eqnarray}}
\newcommand{\eea}{\end{eqnarray}}

\newcommand\reallywidehat[1]{\arraycolsep=0pt\relax%
\begin{array}{c}
\stretchto{
  \scaleto{
    \scalerel*[\widthof{\ensuremath{#1}}]{\kern-.5pt\bigwedge\kern-.5pt}
    {\rule[-\textheight/2]{1ex}{\textheight}} 
  }{\textheight} %
}{0.5ex}\\           
#1\\                 
\rule{-1ex}{0ex}
\end{array}
}

\begin{document}

\title[]{Determinant formula for the field form factor in the anyonic Lieb-Liniger model}
\author{Lorenzo Piroli$^{1,2}$, Stefano Scopa$^{3}$, Pasquale Calabrese$^{3,4}$}
\address{$^1$ Max-Planck-Institut f\"ur Quantenoptik, Hans-Kopfermann-Str. 1, 85748 Garching, Germany}
\address{$^2$ Munich Center for Quantum Science and Technology, Schellingstraße 4, 80799 M\"unchen, Germany}
\address{$^3$ SISSA and INFN, via Bonomea 265, 34136 Trieste, Italy}
\address{$^4$ International  Centre  for  Theoretical  Physics  (ICTP),  I-34151,  Trieste,  Italy}
\date{\today}

\begin{abstract}
We derive an exact formula for the field form factor in the anyonic Lieb-Liniger model, valid for arbitrary values of the interaction $c$, anyonic parameter $\kappa$, and number of particles $N$. 
Analogously to the bosonic case, the form factor is expressed in terms of the determinant of a $N\times N$ matrix, whose elements are rational functions of the Bethe quasimomenta but explicitly depend on $\kappa$. The formula is efficient to evaluate, and provide an essential ingredient for several numerical and analytical calculations. Its derivation consists of three steps. First, we show that the anyonic form factor is equal to the bosonic one between two special \emph{off-shell} Bethe states, in the standard Lieb-Liniger model. Second, we characterize its analytic properties and provide a set of conditions that uniquely specify it. Finally, we show that our determinant formula satisfies these conditions.

\end{abstract}


\maketitle

\section{Introduction}
\label{sec:intro}

Dimensionality plays a crucial role in the study of many-body quantum physics. For instance, the well-known Fermi liquid theory breaks down in one-dimension, due to the drastic effects of interactions compared to the higher dimensional case~\cite{giamarchi_book}. Even more fundamentally, it is closely related to the quantum statistics of indistinguishable particles: while in three spatial dimensions they can only be Bosons or Fermions, two-dimensional systems allow for anyonic statistics, with properties that interpolate between the two~\cite{lm_77} and are responsible for unique physical phenomena, such as, \emph{e.g}., the quantum Hall effect~\cite{laughlin_83}.

As a recent development, a series of studies suggested the possibility of confining anyons in one dimension, by exploiting experimental techniques that are in principle already available within cold atom physics~\cite{klmr-11,gs-15,sse-16}. These works paralleled several theoretical speculations on the subject, which culminated in the study of concrete models of one-dimensional anyons~\cite{agjp-96,rabello-96,it-99,kundu-99,lmp-00,girardeau-06,bgo-06,an-07,pka-07,bgh-07,BeCM09}. Focusing mainly on ground-state and thermal physics, these investigations already provided quantitative predictions for several quantities, including correlation functions \cite{bgh-07,cm-07,pka-10,pka-10_II,sc-09,Patu19} and the (particle) entanglement entropy~\cite{ssc-07,ghc-09}. As a distinguished qualitative feature, it was found that anyonic gases at equilibrium display a momentum distribution that is not symmetric, signaling the fact that the Hamiltonian breaks parity symmetry~\cite{pka-08,sc-08,hzc-08,pka-09b,patu-15,Zinn15,hao-16,mpc-16 }.

The vast majority of these studies was restricted to infinitely repulsive anyons, since finite interactions bring about additional computational challenges. This is true also in the integrable anyonic Lieb-Liniger gas~\cite{pka-07,bgh-07}, which generalizes the well-known model of point-wise interacting Bosons introduced long ago~\cite{ll-63}. Indeed, despite the underlying integrability, the computation of correlation functions is a hard problem. In the bosonic case, after many years of technical advances, beautiful results have been derived for the ground-state~\cite{JiMi81,OlDu03,GaSh03,CaCa06,ChSZ05,ChSZ06,ScFl07,PiCa16-1} and at thermal equilibrium~\cite{MiVT02,KGDS03,SGDV08,KuLa13,PaKl13,ViMi13,PaCa14,NRTG16}. However, so far much less attention has been devoted to the anyonic case.

In the past ten years, the ability to compute correlation functions in integrable systems have become even more urgent, due to the possibility of realizing them using cold atom settings and of testing directly theoretical predictions against experiments~\cite{bdz-08,ccgo-11}. Furthermore, the increasing interest in isolated systems out of equilibrium~\cite{pssv-11,CaEM16} has provided a strong motivation to compute correlation functions for arbitrary excited states. In the Lieb-Liniger gas, this led to the discovery of new analytic formulas for one-point correlation functions~\cite{KoMT09,KoCI11,KoMT11_sTG,Pozs11_mv,Pozs11,PiCE16,BaPi18,BaPC18} and \emph{form factors}~\cite{CaCaSl07,KoMP10,Pozs11,PiCa15} (the first determinant formula for the field form factor in this model was obtained in Ref.~\cite{KoKS97}, see also Refs.~\cite{IzKR87,KoSl99}). The latter are matrix elements of local operators between different energy eigenstates, and are particularly important out of equilibrium. Indeed, they represent one of the building blocks of the so-called quench action method, an analytical approach to tackle exactly the dynamics of interacting integrable systems~\cite{CaEs13,DeCa14,DWBC14,DePC15,Caux16}.

In the case of anyons, several studies have already shown that intriguing properties emerge beyond equilibrium physics, including, for instance, a ``dynamical fermionization'' which appears to be quite robust against different protocols~\cite{del_Campo08,hc-12,Li-15,wrdk-14,PiCa17}. However, these works were restricted, once again, to the infinitely repulsive regime, while up to now no analytical tools were available to tackle the case of finite interactions.

In this paper we study the form factors between arbitrary energy eigenstates in the  anyonic Lieb-Liniger  gas, and derive an exact formula for the creation and annihilation fields of the model. The latter is valid for all the values of anyonic parameter and interaction, both repulsive and attractive, and for any number of particles $N$. Analogously to the bosonic case~\cite{KoKS97, CaCaSl07}, the form factor is expressed in terms of the determinant of a $N\times N$ matrix, whose elements are rational functions of the eigenstate quasimomenta, and is efficient to evaluate. Our formula provides an important first step towards the study of the model beyond the infinitely repulsive regime, both in and out of equilibrium.

Due to the anyonic fractional statistics, the computation of correlation functions in the anyonic Lieb-Liniger gas presents additional difficulties with respect to the bosonic case, which can be studied by means of standard tools within the Algebraic Bethe Ansatz~\cite{korepinBook} (ABA). For this reason, instead of tackling the computation directly in the anyonic model, we map the form factor to the matrix element of the \emph{bosonic} field between two special \emph{off-shell} Bethe states (which will be defined in the next section). This allows us to make use of standard techniques within the ABA formalism, and follow the strategy developed in Ref.~\cite{PiCa15}, where a set of determinant formulas were derived in the bosonic case.

The rest of this article is organized as follows. In Sec.~\ref{sec:model} we introduce the anyonic Lieb-Liniger model, and its solution using the Bethe Ansatz. In Sec.~\ref{sec:summary_result} we present our main result, namely the determinant formula for the field form factor, which is derived in the rest of the paper. In particular, in Sec.~\ref{sec:mapping_to_bosons} we map the problem to a computation of matrix elements in the bosonic Lieb-Liniger model. This is then tackled using the Algebraic Bethe Ansatz, which is reviewed for convenience in Sec.~\ref{sec:bethe_ansatz}. In Sec.~\ref{sec:computation_of_norms} we derive a formula for the norm, while the one for the form factors is finally proven in Sec.~\ref{sec:computation_of_ff}. Our conclusions are consigned to Sec.~\ref{sec:from_aba_to_wave}, while the most technical aspects of our work are reported in a few appendices.

\section{The model and the field form factors}
\label{sec:model}

\subsection{The Hamiltonian and the Bethe Ansatz solution}

We consider the anyonic Lieb-Liniger Hamiltonian~\cite{bgh-07}
\be
H_{LL}=\int_{0}^{L}  {d} x\left[\partial_{x} \Psi_{A}^{\dagger}(x)\partial_{x} \Psi_{A}(x)+c \Psi_{A}^{\dagger}(x) \Psi_{A}^{\dagger}(x) \Psi_{A}(x) \Psi_{A}(x)\right]\,,
\label{eq:hamiltonian}
\ee
where $c$ is the coupling constant of the model, which describes a gas of anyons confined in the segment $[0,L]$. The anyonic fields satisfy the commutation relations
\bea
\Psi_{A}\left(x_{1}\right) \Psi_{A}^{\dagger}\left(x_{2}\right)&=& e^{- {i} \pi \kappa \epsilon\left(x_{1}-x_{2}\right)} \Psi_{A}^{\dagger}\left(x_{2}\right) \Psi_{A}\left(x_{1}\right)+\delta\left(x_{1}-x_{2}\right)\,, \label{eq:commutation_field_1}\\ 
\Psi_{A}^{\dagger}\left(x_{1}\right) \Psi_{A}^{\dagger}\left(x_{2}\right)&=& e^{ {i} \pi \kappa \epsilon\left(x_{1}-x_{2}\right)} \Psi_{A}^{\dagger}\left(x_{2}\right) \Psi_{A}^{\dagger}\left(x_{1}\right)\,, \label{eq:commutation_field_1I}\\
\Psi_{A}\left(x_{1}\right) \Psi_{A}\left(x_{2}\right)&=& e^{ i \pi \kappa \epsilon\left(x_{1}-x_{2}\right)} \Psi_{A}\left(x_{2}\right) \Psi_{A}\left(x_{1}\right)\,,\label{eq:commutation_field_1II}
\eea
where
\be
\epsilon(x)=\left\{
\begin{array}{cc}
	+1\,, & x>0\,,\\
	-1\,, & x<0\,,\\
	0\,, & x=0\,.
\end{array}\right.
\label{eq:epsilon_function}
\ee
Here $\kappa$ is the statistics parameter, and the above expressions reduce to the traditional bosonic and fermionic commutation relations for $\kappa=0, 1$ respectively.

The Hamiltonian \eqref{eq:hamiltonian} generalizes the well known bosonic Lieb-Liniger model~\cite{ll-63}. It was introduced and solved using the Bethe Ansatz by Kundu \cite{kundu-99}, and systematically analyzed by Batchelor \emph{ et al.} \cite{bgo-06,bgh-07} and P\^atu \emph{et al.} \cite{pka-07,pka-10,pka-10_II}. In the following, we briefly review the aspects of its solution that are directly relevant for our purposes. Note that we will employ the same conventions used in Ref.~\cite{pka-10}.

We start by recalling that the Hamiltonian~\eqref{eq:hamiltonian} acts on states of the form
\be
\fl \ket{\Psi_{N}}=\frac{1}{\sqrt{N !}} \int_{0}^{L}  {d} z_{1} \cdots \int_{0}^{L}  {d} z_{N} \chi_{N}\left(z_{1}, \ldots, z_{N} \right) \Psi_A^{\dagger}\left(z_{N}\right) \cdots \Psi_A^{\dagger}\left(z_{1}\right)|0\rangle\,,
\label{eq:states}
\ee
where $|0\rangle$ is the Fock vacuum state. For consistency with Eqs.~\eqref{eq:commutation_field_1}--\eqref{eq:commutation_field_1II}, the wave function has to satisfy
\be
\chi_{N}\left(\ldots, z_{i}, z_{i+1}, \ldots\right)= {e}^{ {i} \pi \kappa \epsilon\left(z_{i}-z_{i+1}\right)} \chi_{N}\left(\ldots, z_{i+1}, z_{i}, \ldots\right)\,.
\label{eq:comm_relation}
\ee
Note that the order of the field operators in Eq.~\eqref{eq:states} is important. 

Next, the eigenvalue equation $H_{LL}|\psi\rangle_{N}=E_{N}|\psi\rangle_{N}$ can be reduced to the quantum-mechanical problem~\cite{bgo-06,pka-07}
\be
\mathcal{H}_N \chi_{N}\left(z_{1}, \ldots, z_{N} \right) =E_{N} \chi_{N}\left(z_{1}, \ldots, z_{N} \right) \,,
\label{eq:quantum_eigen}
\ee
where
\be
\mathcal{H}_{N}=\sum_{j=1}^{N}\left(-\frac{\partial^{2}}{\partial z_{j}^{2}}\right)+2 c \sum_{1 \leqslant j \leqslant k \leqslant N} \delta\left(z_{j}-z_{k}\right)\,.
\ee
Eq.~\eqref{eq:quantum_eigen} has to be supplemented with a set of boundary conditions for the quantum mechanical wave function $\chi_N$. As discussed in \cite{pka-07}, the anyonic commutation relations are not consistent with requiring periodic boundary conditions for all their coordinates $z_j$. In this work, following Ref.~\cite{pka-10}, we will make the consistent choice
\bea
\chi_{N}\left(0, z_{2}, \ldots, z_{N}\right)&=& \chi_{N}\left(L, z_{2}, \ldots, z_{N}\right)\,, \label{eq:BC_a}\\ 
\chi_{N}\left(z_{1},0, \ldots, z_{N}\right)&=& {e}^{ {i}2 \pi \kappa} \chi_{N}\left(z_{1}, L, \ldots, z_{N}\right)\,, \\
&\vdots & \nonumber \\ 
\chi_{N}\left(z_{1}, z_{2}, \ldots,0\right)&=& {e}^{ {i}2 \pi(N-1) \kappa} \chi_{N}\left(z_{1}, z_{2} \ldots, L\right)\,.\label{eq:BC_z}
\eea

The eigenvalue problem~\eqref{eq:quantum_eigen}, with the additional conditions~\eqref{eq:comm_relation} and~\eqref{eq:BC_a}--~\eqref{eq:BC_z} was solved in Ref.~\cite{kundu-99,bgo-06,pka-07} using the Coordinate Bethe Ansatz approach. As in the well-known bosonic case~\cite{ll-63}, it was found that each $N$-particle eigenstate is associated with a set of quasimomenta, or  \emph{rapidities} $\{\lambda_j\}_{j=1}^N$ which generalize the concept of particle momenta for free Fermi gases. The rapidities $\lambda_j$ must satisfy the Bethe equations
\be
e^{ i \lambda_{j} L}= e^{ -i \pi \kappa (N-1)} \prod_{k=1\atop  k \neq j}^{N}\left(\frac{\lambda_{j}-\lambda_{k}+ {i} c^{\prime}}{\lambda_{j}-\lambda_{k}- {i} c^{\prime}}\right)\,,
\label{eq:bethe_equations}
\ee
where
\be
c^\prime=\frac{c}{\cos(\kappa \pi/2)}\,.
\ee
Given a solution to the Bethe equations~ \eqref{eq:bethe_equations}, we can write, up to an arbitrary normalization, the wave function of the corresponding  eigenstate as~\cite{pka-07}
\bea
\chi_{N}\left(z_{1}, \ldots, z_{N} |\{\lambda_j\}\right)&=&\frac{ c^{\prime N/2}}{\sqrt{N ! } } e^{+ {i} \frac{\pi \kappa}{2} \sum_{j<k} \epsilon\left(z_{j}-z_{k}\right)}\nonumber\\
&\times& \sum_{\pi \in S_{N}}e^{ {i} \sum_{n=1}^{N} z_{n} \lambda_{\pi(n)}} \prod_{j<k}\left[1- \frac{{i} c^{\prime} \epsilon\left(z_{k}-z_{j}\right)}{\lambda_{\pi(k)}-\lambda_{\pi(j)}}\right]\,,
\label{eq:wave_function}
\eea
where $\epsilon(x)$ is defined in Eq.~\eqref{eq:epsilon_function}, while the sum is over all the permutations $\pi\in S_N$ of the $N$ rapidities. 
The corresponding energy eigenvalue is
\bea
E\left[\{\lambda_j\}_{j=1}^N\right]=\sum_{j=1}^{N}\lambda_j^{2}\,.\label{eq:finite_size_energy}
\eea
The wave function~\eqref{eq:wave_function} defines a state also when the rapidities do not satisfy the Bethe  equations, which we call off-shell. In the case when the Bethe equations are instead satisfied, we call the state corresponding to~\eqref{eq:wave_function} on-shell.

\subsection{The field form factors}

In this work we are interested in the form factors of the  creation and annihilation operators. Explicitly, we consider
\bea
\fl F_{N+1, N}\left[x; \{\lambda_j\}_{j=1}^{N+1},\{\mu_j\}_{j=1}^{N}\right]=\frac{1}{\sqrt{(N+1) ! N !}} \int {d}^{N+1} y {d}^{N} z \chi_{N+1}^{*}\left(y_{1}, \ldots, y_{N+1} |\{\lambda\}\right) \nonumber \\
\fl\times \chi_{N}\left(z_{1}, \ldots, z_{N} |\{\mu\}\right)   \left\langle 0\left|\Psi_{A}\left(y_{1}\right) \cdots \Psi_{A}\left(y_{N+1}\right) \Psi_{A}^{\dagger}(x) \Psi_{A}^{\dagger}\left(z_{N}\right) \cdots \Psi_{A}^{\dagger}\left(z_{1}\right)\right| 0\right\rangle\,,
\eea
and
\bea
\fl G_{N, N+1}\left[x; \{\mu_j\}_{j=1}^{N},\{\lambda_j\}_{j=1}^{N+1}\right]=\frac{1}{\sqrt{(N+1) ! N !}} \int {d}^{N} y {d}^{N+1} z \chi_{N}^{*}\left(y_{1}, \ldots, y_{N} |\{\mu_j\}\right) \nonumber\\
\fl \times \chi_{N+1}\left(z_{1}, \ldots, z_{N+1} |\{\lambda_j\}\right)
 \left\langle 0\left|\Psi_{A}\left(y_{1}\right) \cdots \Psi_{A}\left(y_{N}\right) \Psi_{A}(x) \Psi_{A}^{\dagger}\left(z_{N+1}\right) \cdots \Psi_{A}^{\dagger}\left(z_{1}\right)\right| 0\right\rangle\,.
 \label{eq:ff_g_wave}
\eea
These expressions can be rewritten using the commutation relations~\eqref{eq:commutation_field_1}--\eqref{eq:commutation_field_1II}. In particular, it is straightforward to compute~\cite{pka-07}
\bea
\fl F_{N+1,N}\left[x; \{\lambda_j\}_{j=1}^{N+1},\{\mu_j\}_{j=1}^{N}\right]=\sqrt{N+1} \int {d}^{N} z \chi_{N+1}^{*}\left(z_{1}, \ldots, z_{N },x |\{\lambda\}\right)\nonumber\\
\times \chi_{N}\left(z_{1}, \ldots, z_{N} |\{\mu\}\right)\,,
\label{eq:ff_integral}
\eea
while we simply have
\be
G_{N, N+1}\left[x; \{\mu_j\}_{j=1}^{N},\{\lambda_j\}_{j=1}^{N+1}\right]=\left(F_{N+1\,N}\left[x; \{\lambda_j\}_{j=1}^{N+1},\{\mu_j\}_{j=1}^{N}\right]\right)^\ast\,.
\label{eq:annihilation_ast}
\ee
Note that the wave functions are not normalized. Then, in order to obtain the normalized form factors, we also need to compute
\bea
\braket{\Psi_N|\Psi_N}=\int {d}^{N} z\, \chi_{N}^{*}\left(z_{1}, \ldots, z_{N} |\{\lambda\}\right)\chi_{N}\left(z_{1}, \ldots, z_{N} |\{\lambda\}\right)\,.
\label{eq:norm_wave}
\eea

\section{Summary of our results}
\label{sec:summary_result}

\subsection{The field form factor}

Let $\{\mu_j\}_{j=1}^{N}$ and $\{\lambda_j\}_{j=1}^{N+1}$ be two sets of rapidities satisfying the Bethe equations \eqref{eq:bethe_equations}, such that $\mu_j\neq \lambda_k$, $\forall j,=1,\ldots ,N$, $k=1,\ldots ,N+1$.  
Our main result is the following formula
\bea
\fl G_{N, N+1}\left[x; \{\mu_j\}_{j=1}^{N},\{\lambda_j\}_{j=1}^{N+1}\right]=-\frac{i}{\sqrt{c^\prime}}\exp\left[i\left(\sum_{j=1}^{N+1}\lambda_j-\sum_{j=1}^{N}\mu_j\right)x\right]\prod_{j,k=1}^{N+1}\left(\lambda_{jk}+ic^\prime\right)\nonumber\\
\times\left(\prod_{j=1}^{N+1}\prod_{k=1}^{N}\frac{1}{\lambda_j-\mu_k}\right)\prod_{j=1}^{N+1}\left(V^{+}_{j}(\kappa)-V^{-}_{j}(\kappa)\right)\frac{\det_{N+1}\left(\delta_{jk}+U_{jk}\right)}{\left(V^{+}_{p}(\kappa)-V^{-}_{p}(\kappa)\right)}\,, 
\label{eq:ff_formula}
\eea
where $\lambda_{jk}=\lambda_{j}-\lambda_{k}$, while
\bea
V^{\pm}_{j}(\kappa)=e^{\mp i \pi\kappa/2}\frac{\prod_{m=1}^{N}\mu_m-\lambda_j \pm ic^\prime}{\prod_{m=1}^{N+1}\lambda_m-\lambda_j \pm ic^\prime}\,,\label{eq:wvpm}\\
U_{jk}=\frac{i}{V^+_{j}(\kappa)-V^{-}_{j}(\kappa)}\frac{\prod_{m=1}^N(\mu_{m}-\lambda_j)}{\prod_{m=1\atop m\neq j}^{N+1}(\lambda_m-\lambda_j)}\left[\mathcal{Q}_{\kappa}(\lambda_j,\lambda_k)-\mathcal{Q}_{\kappa}(\lambda_p,\lambda_k)\right]\,,\label{eq:u_r_matrix}\\
\mathcal{Q}_{\kappa}(x,y)=\cos(\pi\kappa/2)K^{+}(x,y)+i\sin(\pi\kappa/2)K^{-}(x,y)\,,
\label{eq:q_function}
\eea
with
\bea
K^{+}(x,y)=\frac{2c^\prime}{(x-y)^2+c^{\prime 2}}\,,\label{eq:k_plus}\\
K^{-}(x,y)=\frac{2i(x-y)}{(x-y)^2+c^{\prime 2}}\,.\label{eq:k_minus}
\eea
Finally, $\lambda_p$ is an arbitrary complex constant (the value of Eq.~\eqref{eq:ff_formula} does not depend on $\lambda_p$). 

\subsection{The norm}

As a byproduct of our study,  we also obtained a formula for the norm of on-shell Bethe states, which is of course crucial in order to compute normalized form factors. It turns out that it is expressed in terms of the same Gaudin matrix appearing in the Lieb-Liniger model (after substituting $c\to c^\prime$) . Explicitly, we derived
\be
\braket{\Psi_N|\Psi_N}=c^{\prime N}\prod_{j<k}\frac{(\lambda_j-\lambda_k)^2+c^{\prime 2}}{(\lambda_j-\lambda_k)^2}\det \mathcal{G}_{j,k}\,,
\label{eq:norm_eq}
\ee
where
\be
\mathcal{G}_{j,k}=\left[L+\sum_{r=1}^N K^+(\lambda_j,\lambda_r)\right]\delta_{j,k}-K^+(\lambda_j,\lambda_k)\,.
\label{eq:gaudin_elements}
\ee

\subsection{Numerical checks and discussions}

We have tested the validity of Eqs.~\eqref{eq:ff_formula} and~\eqref{eq:norm_eq} against direct numerical calculations, for different values of $x$, $c$, $L$, $\kappa$ and $N$. In particular, we have computed both the form factor and the norm for different energy eigenstates by performing numerically the multi-dimensional integrals of the wave functions, for $N=2,3,4,5$ (we mainly focused on the ground-state and small excitations above it). The result obtained in this way was always found to be in agreement, up to the precision of the numerical integration, with our analytic formulas.

Due to the increasing complexity of the Bethe wave functions, we could not test our formulas for higher values of $N$. We note, however, that the test is already highly nontrivial for $N=5$, where the direct calculation involves integration of many terms over a four-dimensional space. We were able to perform such integrals using the program Mathematica, and found that our prediction was always verified with a relative error smaller than ~$\epsilon\sim 10^{-6}$, which we attribute to the inaccuracy of the multidimensional integration (as expected, the relative error was found to decrease for smaller values of $N$). 

Finally, we comment on the fact that Eq. \eqref{eq:ff_formula} is valid only for sets of rapidities $\{\mu_j\}_{j=1}^N$, $\{\lambda_j\}_{j=1}^{N+1}$ with $\mu_j\neq \lambda_k$ $\forall j,k$. A priori, it might happen that for different on-shell Bethe  states $\mu_j=\lambda_k$ for some $j,k$. However, at least in the limit of large $c$ (and $\kappa\neq 1$) it was shown in Ref.~\cite{pka-10} that this can not happen. For finite values of $c$ and $\kappa$, in analogy with the bosonic case~\cite{CaCaSl07},  we still expect this not to happen except possibly  for a negligibly small number of states, due to the strong constraints imposed by the Bethe equations~\eqref{eq:bethe_equations}.

\section{Preliminary observations and mapping to a bosonic form factor}
\label{sec:mapping_to_bosons}

In this section we show how the anyonic form factor between on-shell Bethe states can be mapped onto the matrix element of the Bose field between special off-shell Bethe states in the standard Lieb-Liniger model.  The starting point is given by the following formula, relating the form factors of the field at different points in space
\bea
F_{N+1,N}\left[x; \{\lambda_j\}_{j=1}^{N+1},\{\mu_j\}_{j=1}^{N}\right]&=&\exp\left[-i\left(\sum_{j=1}^{N+1}\lambda_j-\sum_{j=1}^{N}\mu_j\right)x\right] \nonumber\\
&&\times F_{N+1,N}\left[0; \{\lambda_j\}_{j=1}^{N+1},\{\mu_j\}_{j=1}^{N}\right]\,.
\label{eq:x_dependence}
\eea
This is the same relation that holds in the bosonic Lieb-Liniger model, and that was also derived in Ref.~\cite{pka-10} for the case of infinitely repulsive anyons. In fact, it is possible to prove it also in the case of finite interactions, based exclusively on the form of the Bethe wave functions and on the fact that the rapidities satisfy the Bethe equations~\eqref{eq:bethe_equations}. In particular, one can show that
\bea
\frac{d}{dx}F_{N+1,N}\left[x; \{\lambda_j\}_{j=1}^{N+1},\{\mu_j\}_{j=1}^{N}\right]&=&-i\left(\sum_{j=1}^{N+1}\lambda_j-\sum_{j=1}^{N}\mu_j\right)\nonumber\\
&&\times F_{N+1,N}\left[x; \{\lambda_j\}_{j=1}^{N+1},\{\mu_j\}_{j=1}^{N}\right]\,,\label{eq:der_to_prove}
\eea
which yields immediately  Eq.~\eqref{eq:x_dependence}. The derivation of Eq.~\eqref{eq:der_to_prove} is straightforward, but involves very unwieldy manipulations of the Bethe wave functions. For this reason, we report the proof of Eq.~\eqref{eq:der_to_prove} in~\ref{sec:x_dependence}. Note that taking complex conjugation of Eq.~\eqref{eq:x_dependence} we also obtain
\bea
G_{N,N+1}\left[x; \{\mu_j\}_{j=1}^{N},\{\lambda_j\}_{j=1}^{N+1}\right]&=&\exp\left[i\left(\sum_{j=1}^{N+1}\lambda_j-\sum_{j=1}^{N}\mu_j\right)x\right]\nonumber\\
&&\times G_{N,N+1}\left[0; \{\mu_j\}_{j=1}^{N},\{\lambda_j\}_{j=1}^{N+1}\right]\,.
\label{eq:x_dependence_annihilation}
\eea

Eq.~\eqref{eq:x_dependence} is extremely useful, as it allows us to only focus on the computation of the form factor at $x=0$. Dropping the dependence on the rapidities, we can rewrite it in terms of the Bethe wave functions as
\bea
\fl F_{N+1,N}(0)=\frac{c^{\prime (2N+1)/2}}{N!} e^{-i  \pi\kappa N/2}\sum_{\sigma \in S_{N+1}}\sum_{\pi \in S_{N}}\int d^N z\,  e^{-i\sum_{j=1}^N z_j\left(\lambda_{\sigma(j)}-\mu_{\pi(j)}\right)}\nonumber\\
\fl \qquad \times  \prod_{j=1}^N\left[1 - \frac{ic^\prime }{\lambda_{\sigma(N+1),\sigma(j)}}\right] \left(\prod_{N\geq j>k\geq 1}\left[1+\frac{ic^\prime \epsilon(z_j-z_k)}{\lambda_{\sigma(j),\sigma(k)}}\right]\left[1-\frac{ic^\prime \epsilon(z_j-z_k)}{\mu_{\sigma(j),\sigma(k)}}\right]\right),
\label{eq:integral_expression}
\eea
where we employed the shorthand notation $\lambda_{jk}=\lambda_{j}-\lambda_{k}$, and used that $z_j\geq 0$. 

Up to the trivial prefactor $e^{ - i  \pi\kappa N/2}$, the r.h.s. of Eq.~\eqref{eq:integral_expression} is almost the same expression for the form factor that one would obtain in the bosonic Lieb-Liniger model. There are, however, two differences. First, the interaction $c$ has to be replaced by its effective value $c^\prime$. Second, and most importantly, the rapidities $\lambda_j$ and $
\mu_j$ must satisfy the anyonic, and not the bosonic, Bethe equations~\eqref{eq:bethe_equations}, with a nonzero statistics parameter $\kappa$. 

In summary, we showed that in order to obtain the anyonic form factors, we can compute them in the bosonic Lieb-Liniger model (with interaction set to $c^\prime$) between two off-shell Bethe states, whose rapidities satisfy a set of ``twisted'' boundary conditions given by Eq.~\eqref{eq:bethe_equations}. We note that the twist in the Bethe equations depends on the number of rapidities, which is an essential ingredient determining the nontrivial features of the final result, and which is different from what encountered, e.g., in integrable spin chains with diagonal twisted boundary conditions.

In order  to compute these form factors in the Lieb-Liniger model, we will follow the same strategy developed in Ref.~\cite{PiCa15}, where determinant formulas for several observables were derived. On the technical level, this is based on the Algebraic Bethe Ansatz approach; in the next section, we briefly review it, and present the aspects that are needed in our subsequent derivations.

\section{The Algebraic Bethe Ansatz}
\label{sec:bethe_ansatz}

In this section we introduce the Algebraic Bethe Ansatz, which is the natural framework for the computation 
of correlation functions in integrable models~\cite{korepinBook}. In particular, here we describe the ABA for the (bosonic) Lieb-Liniger model, with interaction strength $c^\prime$. 

One of the fundamental objects in this framework is the monodromy matrix
\begin{equation}
T(\lambda)= \left(\begin{array}{cc}
A(\lambda)&B(\lambda)\\
C(\lambda)&D(\lambda)
\end{array}\right)\ ,
\label{eq:monodromy}
\end{equation}
where $A(\lambda)$, $B(\lambda)$, $C(\lambda)$, $D(\lambda)$ are operators acting on a reference state which we denote by $|0\rangle$. These operators satisfy a set of nontrivial commutation relations encoded in the famous Yang-Baxter equations, involving the  $R$-matrix
\begin{equation}
R(\lambda,\mu)=
\left(\begin{array}{cccc}
f(\mu,\lambda)& & & \\
&g(\mu,\lambda)&1 &\\
& 1 & g(\mu,\lambda) & \\
& & & f(\mu,\lambda)
\end{array}\right)\ ,
\label{eq:r_matrix}
\end{equation}
where the empty entries are defined to be vanishing and where
\bea
f(\lambda,\mu)&=&\frac{\lambda-\mu+ic^\prime}{\lambda-\mu}\,, \label{eq:f_function}\\
g(\lambda,\mu)&=&\frac{i c^\prime}{\lambda -\mu}\,.\label{eq:g_function}
\eea
The operator entries of the monodromy matrix act on the Hilbert space generated by the Bethe states \begin{equation}
\prod_{j=1}^{N}B(\lambda_j)|0 \rangle\ ,
\label{eq:bethe_states}
\end{equation}
with dual states
\begin{equation}
\langle 0|\prod_{j=1}^{N}C(\lambda_j)\ ,
\end{equation}
while the action on the reference state of the operators $A(\lambda)$, $D(\lambda)$ is given by
\begin{equation}
A(\lambda)|0\rangle=a(\lambda)|0\rangle\ , \qquad D(\lambda)|0\rangle=d(\lambda)|0\rangle.
\label{eq:a_d_operators}
\end{equation}
In the Lieb-Liniger model the functions $a(\lambda)$, $d(\lambda)$ are 
\begin{equation}
a(\lambda)=e^{-i\frac{L}{2}\lambda}\ ,\qquad d(\lambda)=e^{i\frac{L}{2}\lambda}\ .
\label{eq:a_d_functions}
\end{equation}
We further define for later convenience the function
\begin{equation}
r(\lambda)=\frac{a(\lambda)}{d(\lambda)}=e^{-i\lambda L}\ .
\label{eq:r_function}
\end{equation}
Finally, it is useful to define rescaled operators
\begin{equation}
\mathcal{B}(\lambda)=\frac{1}{d(\lambda)}B(\lambda)\ ,\qquad \mathcal{C}(\lambda)=\frac{1}{d(\lambda)}C(\lambda)\ ,
\label{eq:b_c_operators}
\end{equation}
where $d(\lambda)$ is given in \eqref{eq:a_d_functions}.

In the following we will be interested in two quantities. The first one is the norm
\be
\mathcal{N}\left[\{\lambda_j\}_j\right]=\lim_{\{\mu_j\}_j\to \{\lambda_j\}_j}\langle 0| \mathcal{C}(\lambda_1)\cdots \mathcal{C}(\lambda_N)\mathcal{B}(\mu_N)\cdots \mathcal{B}(\mu_1)|0\rangle\,,
\label{eq:norm_general}
\ee
while the second one is the field form factor
\be 
\fl \mathcal{G}_{N}(\{\mu_j\}_{j=1}^N,\{\lambda_j\}_{j=1}^{N+1},\{r(\mu_j)\}_{j=1}^N,\{r(\lambda_j)\}^{N+1})=\langle 0|\prod_{j=1}^{N}\mathcal{C}(\mu_j)\Psi(0)\prod_{j=1}^{N+1}\mathcal{B}(\lambda_j)|0\rangle\ .
\label{eq:MathG_function}
\ee
In both Eqs.~\eqref{eq:norm_general} and \eqref{eq:MathG_function} we made use of the fact that $B^{\dagger}(\bar{\mu})=C(\mu)$, where $\bar{\mu}$ is the complex conjugated of $\mu$~\cite{Kore82}.

The framework summarized in this section can be considered as the algebraic counterpart of the wave-function formalism of the Coordinate Bethe Ansatz. In particular, all the quantities computed using the representation~\eqref{eq:bethe_states} for the Bethe states can be related to those obtained using the wave functions~\eqref{eq:wave_function} (for $\kappa=0$). For instance, it can be shown that the function $\mathcal{G}_{N}$ in Eq.~\eqref{eq:MathG_function} is proportional to $G_{N,N+1}$ defined in Eq.~\eqref{eq:ff_g_wave}, which is what we aim to compute.  This will be shown in detail in Sec.~\ref{sec:from_aba_to_wave}. In the following two sections, instead, we will focus entirely on the algebraic formalism and derive determinant formulas for $\mathcal{N}$ and $\mathcal{G}_N$ defined above.

\section{Computation of the norms}
\label{sec:computation_of_norms}

In this section we provide a determinant formula for the norm~\eqref{eq:norm_general}. Let us first define the scalar product between different Bethe states
\be
\fl \mathcal{S}\left[\{\lambda_j\}_{j=1}^{N},\{\mu_j\}_{j=1}^{N},\{r(\lambda_j)\}_{j=1}^{N}, \{r(\mu_j)\}_{j=1}^N\right]=\langle 0| \mathcal{C}(\lambda_1)\cdots \mathcal{C}(\lambda_N)\mathcal{B}(\mu_N)\cdots \mathcal{B}(\mu_1)|0\rangle\,.
\label{eq:scalar}
\ee
This is a well-studied object in the ABA formalism~\cite{Kore82,korepinBook}. Using the known commutation relations between the matrix elements of the monodromy matrix~\eqref{eq:monodromy} and Eq.~\eqref{eq:a_d_operators}, one immediately sees that \eqref{eq:scalar} can be expressed as a rational function of the rapidities, where $r(\lambda_j)$ and $r(\mu_j)$ play the role of functional parameters. In particular, the scalar product can be formally rewritten as~\cite{korepinBook}
\be
\fl \mathcal{S}\left[\{\lambda_j\},\{\mu_j\},\{r(\lambda_j)\}, \{r(\mu_j)\}\right]=\sum\left\{\prod_{j=1}^Nr(\nu^{A}_j)\right\}K\left(
\begin{array}{ll}
	\{\lambda_j\}& \{\mu_j\}\\
	\{\nu_j^A\}& \{\nu_j^B\}
\end{array}
\right)\,.
\label{eq:scalar_general_rational}
\ee
Here the sum is with respect to all partitions of the set $\{\lambda_j\}\cup \{\mu_j\}$ (consisting of $2N$ elements) into two disjoint sets $\{\nu^{A}_j\}$ and $\{\nu^{B}_j\}$ possessing equal numbers of elements; the coefficients $K_N$ are rational functions of the rapidities that do not depend on the vacuum eigenvalues $r(\lambda_j)$, $r(\mu_j)$.

 In the case when the rapidities $\{\lambda_j\}$ satisfy the standard Bethe equations (namely Eqs.~\eqref{eq:bethe_equations} with $\kappa=0$) this rational function  can be expressed in terms of the famous Slavnov formula~\cite{Slav89}. It turns out that the same formula holds, with minor modifications, also when $\{\lambda_j\}$ satisfy the twisted equations~\eqref{eq:bethe_equations}, as we now show.

In order to see this, consider the normalized scalar product
\bea
\fl  \widetilde{\mathcal{S}}\left[\{\lambda_j\},\{\mu_j\},\{r(\lambda_j)\}, \{r(\mu_j)\}\right]
=e^{-i\pi \kappa (N-1)N }\mathcal{S}\left[\{\lambda_j\},\{\mu_j\},\{r(\lambda_j)\}, \{r(\mu_j)\}\right]\,.
\label{eq:tilde_s_I}
\eea
From Eq.~\eqref{eq:scalar_general_rational} it is immediate to see
\bea
\fl \widetilde{\mathcal{S}}\left[\{\lambda_j\},\{\mu_j\},\{r(\lambda_j)\}, \{r(\mu_j)\}\right]
=\mathcal{S}\left[\{\lambda_j\},\{\mu_j\},\{\tilde{r}(\lambda_j)\}, \{\tilde{r}(\mu_j)\}\right]\,,
\label{eq:tilde_s_II}
\eea
where $\tilde{r}(\lambda)= e^{-i\pi \kappa (N-1)} r(\lambda)$. Due to the Bethe equations~\eqref{eq:bethe_equations} we have 
\be
\tilde{r}(\lambda_j)=\prod_{k=1\atop  k \neq j}^{N}\left(\frac{\lambda_{j}-\lambda_{k}-i c^{\prime}}{\lambda_{j}-\lambda_{k}+ i c^{\prime}}\right)\,.
\label{eq:bosonic_r}
\ee

From Eqs.~\eqref{eq:tilde_s_II} and \eqref{eq:bosonic_r} we see that $\widetilde{\mathcal{S}}$ is expressed as the same rational function corresponding to the overlap between an on-shell Bethe state in the bosonic Lieb-Liniger model and an off-shell state with vacuum expectation values $\tilde{r}(\mu_j)$.  We can thus apply directly Slavnov formula~\cite{Slav89} to obtain
\bea
\fl \widetilde{\mathcal{S}}\left[\{\lambda_j\},\{\mu_j\},\{r(\lambda_j)\}, \{r(\mu_j)\}\right] =\left[\prod_{j<k}g(\lambda_j,\lambda_k)g(\mu_k,\mu_j)\right]\prod_{j,k=1}^Nh(\lambda_j,\mu_k) \mathrm{det}_N M_{l,k}\,,
\label{eq:slavnov_anyons}
\eea
where
\begin{equation}
h(\lambda,\mu)=\frac{f(\lambda,\mu)}{g(\lambda,\mu)}\,,
\end{equation}
and
\begin{equation}
M_{l,k}=\frac{g(\lambda_k,\mu_l)}{h(\lambda_k,\mu_l)}-\tilde{r}(\mu_l)\frac{g(\mu_l,\lambda_k)}{h(\mu_l,\lambda_k)}\prod_{m=1}^N\frac{f(\mu_l,\lambda_m)}{f(\lambda_m,\mu_l)}\,.
\end{equation}
From Eq.~\eqref{eq:slavnov_anyons} we can finally compute the norm taking the limit $\{\mu_j\}\to \{\lambda_j\}$. The calculation is completely analogous to the one originally reported in Ref.~\cite{Slav89}, and presents no difficulty. In particular, we have
\bea
 \mathcal{N}&=&\lim_{\{\mu_j\}\to \{\lambda_j\}}\mathcal{S}\left[\{\lambda_j\},\{\mu_j\},\{r(\lambda_j)\}, \{r(\mu_j)\}\right]\nonumber\\
&=&e^{i\pi\kappa (N-1)N}\lim_{\{\mu_j\}_j\to \{\lambda_j\}_j}\widetilde{\mathcal{S}}\left[\{\lambda_j\},\{\mu_j\},\{r(\lambda_j)\}, \{r(\mu_j)\}\right]\nonumber\\
&=&e^{i\pi \kappa (N-1)N }c^{\prime N}\prod_{j<k}\frac{(\lambda_j-\lambda_k)^2+c^{\prime 2}}{(\lambda_j-\lambda_k)^2} \mathrm{det}_{N}\left(\frac{\partial \varphi_j}{\partial \lambda_k}\right)\,,
\label{eq:almost_there_norm}
\eea
where
\bea
\varphi_j&=&i \ln\left\{ \tilde{r}(\lambda_j)\prod_{k\neq j }\frac{f(\lambda_j,\lambda_k)}{f(\lambda_k,\lambda_j)}\right\}\,.
\eea
We recall that here $\tilde{r}(\lambda)=e^{-i\pi\kappa (N-1)}r(\lambda)=e^{-i\pi\kappa (N-1)}e^{-i\lambda L}$. We can simplify Eq.~\eqref{eq:almost_there_norm} further, using
\bea
  \frac{\partial \varphi_j}{\partial \lambda_k}&=&\frac{\partial }{\partial \lambda_k} i \ln\left\{ \tilde{r}(\lambda_j)\prod_{k\neq j }\frac{f(\lambda_j,\lambda_k)}{f(\lambda_k,\lambda_j)}\right\}=\frac{\partial }{\partial \lambda_k}i \ln\left\{ r(\lambda_j)\prod_{k\neq j }\frac{f(\lambda_j,\lambda_k)}{f(\lambda_k,\lambda_j)}\right\}\nonumber\\
&=&\frac{\partial }{\partial \lambda_k}i \ln\left\{ e^{-i\lambda_j L}\prod_{k\neq j }\frac{f(\lambda_j,\lambda_k)}{f(\lambda_k,\lambda_j)}\right\}
\,.
\eea
Then, putting all together, we finally obtain
\be
\mathcal{N}[\left\{\lambda_j\right\}_j]=e^{i\pi \kappa (N-1)N }c^{\prime N}\prod_{j<k}\frac{(\lambda_j-\lambda_k)^2+c^{\prime 2}}{(\lambda_j-\lambda_k)^2} \mathrm{det}_{N}\mathcal{G}_{j,k}\,,
\label{eq:final_partial_norm}
\ee
where the matrix $\mathcal{G}_{j,k}$ is defined in Eq.~\eqref{eq:gaudin_elements}.

\section{Computation of the form factors}
\label{sec:computation_of_ff}

In this section we provide a determinant formula for the field form factor~\eqref{eq:MathG_function} in the case when the two sets of rapidities $\{\lambda_j\}$ and $\{\mu_j\}$ satisfy the twisted Bethe equations~\eqref{eq:bethe_equations}. In fact, we will treat a slightly more general case, as we now explain.

First, analogously the scalar product~\eqref{eq:scalar_general_rational}, the function $\mathcal{G}_N$ in \eqref{eq:MathG_function} can be written as a sum of rational functions of the rapidities, with $r(\lambda_j)$ and $r(\mu_j)$ appearing as functional parameters. In the following, we are interested in the expression that we get by replacing these functional parameters with explicit rational functions. Specifically, introducing 
\begin{equation}
\vartheta_N\left(\{\lambda_k\}_{k=1}^N;x;q\right)= -e^{i q \pi\kappa}\prod_{k=1}^{N}\frac{\lambda_k-x+ic}{\lambda_k-x-ic}\,,
\label{theta_function}
\end{equation}
we choose
\bea
r(\lambda_j)&=&\vartheta_{N+1}(\{\lambda_k\};\lambda_j; q)=e^{i\pi q\kappa}\prod_{k=1\atop k\neq j}^{N+1}\frac{\lambda_k-\lambda_j+ic}{\lambda_k-\lambda_j-ic}\ , \\
r(\mu_j)&=& \vartheta_{N}(\{\mu_k\};\mu_j; q-1)= e^{i\pi (q-1)\kappa}\prod_{k=1\atop k\neq j}^{N}\frac{\mu_k-\mu_j+ic}{\mu_k-\mu_j-ic}\,. 
\eea

With the above definitions,  the expression for the form factor becomes a rational function of the rapidities, with no dependence on the functional parameter $r(\lambda)$ left. With a slight abuse of notation, we continue to denote this function by $\mathcal{G}_N$. More precisely, we define
\begin{eqnarray}
\fl \mathcal{G}_{N}(q, \{\mu_k\},\{\lambda_k\})\nonumber\\
\hspace{-0.5cm} =\mathcal{G}_{N}(\{\mu_k\}_{k=1}^N,\{\lambda_k\}_{k=1}^{N+1}, \{r(\mu_j)\}_{j=1}^N, \{r(\lambda_j)\}_{j=1}^{N+1})\Big|_{\{r(\mu_j)\}=\{\vartheta_N(\{\mu_k\},\ \mu_j,q-1)\} \atop \{r(\lambda_j)\}=\{\vartheta_{N+1}(\{\lambda_k\},\ \lambda_j,q)\}}\,.
\label{eq:on_shell_ff}
\end{eqnarray}
We stress  that this function is defined for arbitrary values of the rapidities and of the parameter $q$, even though it is physically relevant only for sets satisfying the Bethe equations  \eqref{eq:bethe_equations}, and with the choice $q=N$. 

There are arguably different ways to compute the function $\mathcal{G}_{N}(q, \{\mu_k\},\{\lambda_k\})$. In the rest of this section, we will follow the strategy developed in Ref.~\cite{PiCa15} which provides a rather clean way of proving the final result. Note that we will derive a formula which is valid for general $q$, and at the end of the calculation  we can simply set $q=N$. 

\subsection{Analytic properties of the form factor}

Following Ref.~\cite{PiCa15}, we begin by stating two fundamental propositions regarding the form factor $\mathcal{G}_N$. These are presented below, and provide a strict characterization of the latter in terms of its analytic properties.  
\begin{prop}
\label{prop:properties}
Consider the function $\mathcal{G}_{N}$ defined in \eqref{eq:on_shell_ff}. Then the following properties hold
\begin{enumerate}
\item 
\be
\mathcal{G}_{0}(q,\emptyset,\lambda)=(-i\sqrt{c})e^{iq\pi\kappa}\ ;
\label{eq:property_1}
\ee
\item consider $\mu_m\in\{\mu_{j}\}_{j=1}^{N}$; then the asymptotic behavior of $\mathcal{G}_{N}$ as a function of $\mu_m$ is given by
\be
\lim_{\mu_m\to\infty}\mathcal{G}_{N}(q, \{\mu_j\},\{\lambda_j\})=0\,;
\ee
\item consider $\mathcal{G}_{N}(q,\{\mu_j\},\{\lambda_j\})$ as a function of $\mu_m\in\{\mu_j\}_{j=1}^{N}$. Then it is a rational function and its only singularities are first order poles at $\mu_m=\lambda_j$, $j=1,\ldots N+1$;

\item the residues of the form factors are given by the following recursive relations
\bea
\fl \mathcal{G}_{N}(q,\{\mu_j\}_{j=1}^{N},\{\lambda_j\}_{j=1}^{N+1})\Big|_{\mu_m\to\lambda_k}\sim  g(\mu_m,\lambda_k)
\left[e^{i(q-1)\pi\kappa}\prod_{j=1\atop j\neq m}^{N}f(\mu_j,\mu_m)\prod_{j=1\atop j\neq k}^{N+1}f(\lambda_k,\lambda_j)\right.\nonumber\\
\fl \qquad -\left.e^{iq\pi\kappa}\prod_{j=1\atop j\neq k}^{N+1}f(\lambda_j,\lambda_k)\prod_{j=1\atop j\neq m}^{N}f(\mu_m,\mu_j)\right] \mathcal{G}_{N-1}\left(q, \{\mu_j\}_{j\neq m},\{\lambda_j\}_{j\neq k}\right)\, .
\label{eq:recursive}
\eea
\end{enumerate}
\end{prop}

The second proposition tells us that the analytic properties listed above uniquely specify the form factor, and reads as follows.
 
\begin{prop}
	\label{prop:uniqueness}
	Let $\{\mathcal{H}_{N}=\mathcal{H}_{N}(\{\mu_j\}_{j=1}^{N},\{\lambda_j\}_{j=1}^{N+1})\}$ be a family of functions and suppose that $\mathcal{H}_{N}$ satisfies properties $1-4$ of Prop. \ref{prop:properties} for every $N\geq 0$. Then 
	\begin{equation}
		\mathcal{H}_{N}(\{\mu_j\}_{j=1}^{N},\{\lambda_j\}_{j=1}^{N+1})=\mathcal{G}_{N}(q,\{\mu_j\}_{j=1}^{N},\{\lambda_j\}_{j=1}^{N+1})\ .
		\label{eq:thesis}
	\end{equation}
\end{prop}
The proofs of these propositions follow closely the ones reported in Appendices A and B of Ref.~\cite{PiCa15}, with only minor modifications required with respect to the bosonic case. For this reason, we omit them here, and refer the reader to Ref.~\cite{PiCa15} for all the necessary details.

\subsection{The determinant formula}
\label{sec:determinant_formula}

In this section we exhibit a candidate expression for the rational function corresponding to the field form factor. We will then show that it satisfies all the properties of Prop.~\ref{prop:properties}. Thus, thanks to Prop.~\ref{prop:uniqueness}, this will allow us to conclude that it is the correct expression.

The candidate function for the field form factor reads
\begin{eqnarray}
\fl \mathcal{H}_N(\{\mu_j\},\{\lambda_j\})=-\frac{e^{iq\pi \kappa (N+1)}e^{-i\kappa N\pi/2 }}{\sqrt{c^\prime}}\prod_{j,k=1}^{N+1}\left(\lambda_{jk}+ic^\prime\right)\nonumber\\
\times\left(\prod_{j=1}^{N+1}\prod_{k=1}^{N}\frac{1}{\lambda_j-\mu_k}\right)\prod_{j=1}^{N+1}\left(V^{+}_{j}(\kappa)-V^{-}_{j}(\kappa)\right)\frac{\det_{N+1}\left(\delta_{jk}+U_{jk}\right)}{\left(V^{+}_{p}(\kappa)-V^{-}_{p}(\kappa)\right)}\,, 
\label{eq:h_function}
\end{eqnarray}
where $V^{\pm}_{j}$, $U_{jk}$ are given in Eqs.~\eqref{eq:wvpm} and~\eqref{eq:u_r_matrix} respectively, while $\lambda_p$ is an arbitrary complex number.  We have conjectured this expression based on the formula for the bosonic form factor derived in Ref.~\cite{CaCaSl07}. In particular, we looked for the ``minimal modification'' of the latter which could satisfy properties $1$-$4$ of Prop.~\ref{prop:properties}. 

Before proving that Eq.~\eqref{eq:h_function} satisfies the properties of Prop.~\ref{prop:properties}, let us show that $\mathcal{H}_N$ does not depend on the parameters $\lambda_p$. This is done by means of some identities involving sums of rational functions, that are reported for convenience in~\ref{sec:identities}. First, define
\bea
\Xi_j=\frac{\prod_{k=1}^{N}(\mu_k-\lambda_j)}{\prod_{k=1\atop k\neq j}^{N+R}(\lambda_k-\lambda_j)}\,,\\
\Theta_{j}(\kappa)=V^{+}_{j}(\kappa)-V^{-}_{j}(\kappa)\,.
\label{eq:xi}
\eea
For $k=2,\ldots, N+1$ add to the first column of the matrix $\delta_{jk}+U_{jk}$ column $k$ multiplied by $\Xi_{k}/\Xi_{1}$. From identity \eqref{eq:identity_1} in \ref{sec:identities} it follows that the first column becomes proportional to $V_{p}^{+}(\kappa)-V^{-}_{p}(\kappa)$. Exploiting the multilinearity of the determinant we get
\begin{equation}
\frac{\mathrm{det}_{N+1}\left(\delta_{jk}+U_{jk}\right)}{V^{+}_{p}(\kappa)-V^{-}_{p}(\kappa)}=\frac{\mathrm{det}_{N+1}\left(\mathcal{M}_{jk}\right)}{\Xi_{1}}\ ,
\label{eq:aux_a}
\end{equation}
where
\begin{equation}
\mathcal{M}_{jk}=\left\{
\begin{array}{cc}
\frac{\Xi_{j}}{V^+_{j}(\kappa)-V^{-}_{j}(\kappa)}\ , & \mathrm{if\ } k=1\ ,\\
\delta_{jk}+U_{jk}\ , &\mathrm{otherwise\ }.
\end{array}
\right.
\end{equation}
Now, for $k=2,\ldots, N+1$, add to column $k$ of matrix $\mathcal{M}_{jk}$ column $1$ multiplied by $i\mathcal{Q}_{\kappa}(\lambda_p,\lambda_k)$ [defined in Eq.~\eqref{eq:q_function}]. Exploiting again the multilinearity of the determinant we obtain 
\begin{equation}
\mathrm{det}_{N+1}\left(\mathcal{M}_{jk}\right)=\mathrm{det}_{N+1}\left(\widetilde{\mathcal{M}}_{jk}\right)\ ,
\end{equation}
where
\begin{equation}
\widetilde{\mathcal{M}}_{jk}=\left\{
\begin{array}{cc}
\frac{\Xi_{j}}{V^+_{j}(\kappa)-V^{-}_{j}(\kappa)}\ , & \mathrm{if\ } k=1\ ,\\
\delta_{jk}+\frac{i}{V^+_{j}(\kappa)-V^{-}_{j}(\kappa)}\Xi_{j}\mathcal{Q}_\kappa(\lambda_j,\lambda_k)\ , &\mathrm{otherwise\ }.
\end{array}
\right.
\label{eq:aux_b}
\end{equation}
In the final expression \eqref{eq:aux_b} $\lambda_p$ has disappeared: we conclude that the l.h.s. of \eqref{eq:aux_a} and thus $\mathcal{H}_N$ in \eqref{eq:h_function} are independent of the parameter $\lambda_p$. In fact, with this procedure we also see that $\mathcal{H}_N$, as a function of the parameter $\mu_m$, does not have poles corresponding to the zeroes of $V_{p}^{+}(\kappa)-V_{p}^{-}(\kappa)$, since these factors are canceled by the determinant in the numerator in the r.h.s. of Eq. \eqref{eq:h_function}.

We shall now show that $\mathcal{H}_N$ satisfies properties $1-4$ of Prop. \ref{prop:properties}. Property $1$ is trivial. To see that the second is true, it is sufficient to observe that 
\begin{equation}
\mathcal{H}_{N}(\{\mu_j\},\{\lambda_j\})=O(\mu_m^{-1})\,,
\end{equation}
which follows straightforwardly from Eq.~\eqref{eq:h_function}.

Property $3$ is also easy to prove. In  fact, we have already shown  that there are no poles corresponding  to the zeros of $V_{p}^{+}(\kappa)-V_{p}^{-}(\kappa)$, which implies that the only poles can be at $\mu_k=\lambda_j$. 

As the only nontrivial part of the proof, we have to show that property $4$ is satisfied. Suppose $\mu_{r}\to \lambda_{\ell}$.  In this limit, all the elements of the row $\ell$ of the matrix $\delta_{jk}+U_{jk}$ become zero, except for the diagonal one, so it is straightforward to compute
\begin{eqnarray}
\lim_{\mu_r\to\lambda_\ell} \mathcal{H}_N(\{\mu_r\},\{\lambda_\ell\})\sim  g(\mu_r,\lambda_\ell) \left(-\frac{e^{iq\pi \kappa N}e^{-i\kappa (N-1)\pi/2}}{\sqrt{c^\prime }}\right)&\nonumber\\
 \times\left[e^{i(q-1)\kappa\pi}\prod_{j=1\atop j\neq r }f(\mu_j,\mu_r)\prod_{j=1\atop j\neq \ell}f(\lambda_\ell,\lambda_j)-e^{iq\kappa\pi}\prod_{j=1\atop j\neq \ell}f(\lambda_j,\lambda_\ell)\prod_{j=1\atop j\neq r}f(\mu_r,\mu_j)\right] &\nonumber\\
 \times \prod_{j,k=1\atop j,k\neq \ell}\left(\lambda_{jk}+ic^\prime \right)\prod_{j=1\atop j\neq \ell}\prod_{k=1\atop k\neq r}\frac{1}{\lambda_j-\mu_k}\prod_{j=1\atop j\neq \ell}\left(V^{+}_{j}(\kappa)-V^{-}_{j}(\kappa)\right)&\nonumber\\
 \times \frac{1}{\left(V^{+}_{p}(\kappa)-V^{-}_{p}(\kappa)\right)}\mathrm{det}_{N}\left(\delta_{jk}+U_{jk}\right) .&
\label{property_4}
\end{eqnarray}
where $V^{\pm}_{j}$ and $U_{jk}$ are defined in \eqref{eq:wvpm}, \eqref{eq:u_r_matrix} for the sets of rapidities $\{\mu_j\}_{j\neq r}$, $\{\lambda_j\}_{j\neq \ell}$. Comparing with Eq.~\eqref{eq:h_function}, we have thus shown that the function $\mathcal{H}_N$ satisfies property $4$ of Prop.~\ref{prop:properties}.

\section{From the Algebraic Bethe Ansatz to the wave functions }
\label{sec:from_aba_to_wave}

The results derived in the previous sections hold within the framework of the ABA. In this section, we discuss the correct prefactors that must be taken into account when using the wave-function formalism introduced in Sec.~\ref{sec:model}.

The relation between the two frameworks is given by the formula~\cite{korepinBook}
\bea
\fl \langle 0|\Psi(x_N)\cdots \Psi(x_1)\prod_{j=1}^{N}B(\lambda_j)|0\rangle  &=(-i\sqrt{c^\prime})^{ N}\exp\left(-i\frac{L}{2}\sum_{k=1}^N\lambda_k\right)\nonumber\\
&\times \sum_{\mathcal{P}\in S_N}e^{i \sum_{j=1}^N x_j \lambda_{\mathcal{P}_{j}}}\prod_{j<k}\left(1-\frac{i c^\prime\, \epsilon(x_k-x_j)}{\lambda_{\mathcal{P}_k}-\lambda_{\mathcal{P}_j}}\right)\,,
\eea
where $\Psi(x)$ are bosonic operators. This gives us the precise wave function corresponding to the Bethe state~\eqref{eq:bethe_states}. Comparing to the conjugate of Eq.~\eqref{eq:integral_expression}, we conclude
\bea
\langle 0|\prod_{j=1}^{N}C(\mu_j)\Psi(0)\prod_{j=1}^{N+1}B(\lambda_j)|0\rangle=(-i)e^{-i\pi\kappa N/2}\nonumber\\
\quad \times  \exp\left(-i\frac{L}{2}\sum_{k=1}^{N+1}\lambda_k+i\frac{L}{2}\sum_{k=1}^N\mu_k\right)  G_{N,N+1}\left[0; \{\mu_j\}_{j=1}^{N},\{\lambda_j\}_{j=1}^{N+1}\right]\,,
\label{eq:normalization_1}
\eea
and also
\bea
\langle 0|\prod_{j=1}^{N}C(\lambda_j)\prod_{j=1}^{N}B(\lambda_j)|0\rangle=\langle \Psi_N|\Psi_N\rangle\,,
\label{eq:normalization_2}
\eea
where $\langle \Psi_N|\Psi_N\rangle$ is defined in Eq.~\eqref{eq:norm_wave}.
Note that in the previous sections we have always worked with the normalized operators $\mathcal{B}(\lambda)$ and $\mathcal{C}(\lambda)$ defined in Eq.~\eqref{eq:b_c_operators}, which are related to $B(\lambda)$ and $C(\lambda)$ by the prefactor $d(\lambda)=e^{i\lambda L/2}$. Taking this into account, we can derive directly the exact expression for $G_{N,N+1}$ and $\braket{\Psi_N|\Psi_N}$ from Eq.~\eqref{eq:normalization_1} and \eqref{eq:normalization_2}, in terms of the formulas in Eqs.~\eqref{eq:final_partial_norm} and \eqref{eq:h_function}. By doing this, we finally obtain the results anticipated in Eqs.~\eqref{eq:ff_formula} and \eqref{eq:norm_eq}. 

\section{Conclusions}
\label{sec:conclusions}

In this work we have derived an exact formula for the (normalized) field form factor in the anyonic Lieb-Liniger model, valid for arbitrary values of the interaction $c$, anyonic parameter $\kappa$, and number of particles $N$. The final result is a remarkably simple generalization of the bosonic formula first derived in Ref.~\cite{KoKS97} (see also \cite{CaCaSl07}), and is expressed in terms of the determinant of a $N\times N$ matrix whose elements are rational functions of the Bethe rapidities.

 From the physical point of view, our formula represents the starting point for many numerical and analytical calculations. For instance, a natural application of our result would be the computation of the Green function, namely the expectation value of the operator $\Psi^{\dagger}_A(x)\Psi_A(y)$ (also at different times). 
 This could be done very efficiently using the ABACUS algorithm~\cite{mc-05,calabrese_0,caux_2,klauser,PaCa14}, for arbitrary excited states and large numbers of particles.  

 Another particularly interesting direction would be to exploit our formula to derive nonuniversal prefactors in the Luttinger-liquid description of the 1$D$ anyonic gas, along the lines of Refs.~\cite{SGCI11,SPCI12,kms-11}. In turn, this would make it possible to study the system in inhomogeneous settings, both in and out of equilibrium, thus generalizing  recent results obtained in the bosonic Lieb-Liniger model~\cite{BrDu17,RCDD19,bd-18}. We plan to address these problems in future work.

\section*{Acknowledgments}

LP acknowledges support from the Alexander von Humboldt foundation. PC and SS acknowledge support from ERC under Consolidator grant  number
771536 (NEMO).

\appendix

\section{Proof of Eq.~\eqref{eq:der_to_prove}}
\label{sec:x_dependence}	

The proof of Eq.~\eqref{eq:der_to_prove} is nontrivial, and is derived in the following. First, throughout this section we choose a different normalization for  the wave functions, which makes the computations a bit lighter. In particular, we use
\bea
\fl  \chi_{N}(\{z_j\},\{\lambda_j\})= \mathrm{e}^{i \frac{\pi \kappa}{2} \sum_{j<k} \epsilon\left(z_{j}-z_{k}\right)} \nonumber\\
\times \sum_{\pi \in S_{N}}(-1)^{\pi} \mathrm{e}^{\mathrm{i} \sum_{n=1}^{N} z_{n} \lambda_{\pi(n)}} \prod_{j>k}\left[\lambda_{\pi(j)}-\lambda_{\pi(k)}-\mathrm{i} c^{\prime} \epsilon\left(z_{j}-z_{k}\right)\right]\,.
 \eea
Using this wave function, the form factor is written as
\bea
\fl F_{N+1,N}(x)=\sum_{\sigma \in S_{N+1}}\sum_{\pi \in S_{N}}(-1)^{\sigma +\pi}e^{-ix\lambda_{\sigma (N+1)}}\int d^N z\, e^{-i (\pi\kappa /2) \sum_{j=1}^N\epsilon(z_j-x)} e^{-i\sum_{j=1}^N z_j\left(\lambda_{\sigma(j)}-\mu_{\pi(j)}\right)}\nonumber\\
\fl \times  \prod_{j=1}^N\left[\lambda_{\sigma(N+1),\sigma(j)} + ic^\prime \epsilon (x-z_j)\right] \left(\prod_{N\geq j>k\geq 1}\left[\lambda_{\sigma(j),\sigma(k)}+ic^\prime \epsilon(z_j-z_k)\right]\right.\nonumber\\
\left.\left[\mu_{\sigma(j),\sigma(k)}-ic^\prime \epsilon(z_j-z_k)\right]\right),
\eea
where we used the shorthand notation $\lambda_{jk}=\lambda_{j}-\lambda_{k}$. Taking the derivative w.r.t. $x$, we obtain
\bea
\fl \frac{d}{dx}F_{N+1,N}(x)=\sum_{\sigma \in S_{N+1}}\sum_{\pi \in S_{N}}(-1)^{\sigma +\pi}(-i\lambda_{\sigma (N+1)})e^{-ix\lambda_{\sigma (N+1)}}\int d^N z\, e^{-i (\pi\kappa /2) \sum_{j=1}^N\epsilon(z_j-x)} \nonumber\\
\fl \times e^{-i\sum_{j=1}^N z_j\left(\lambda_{\sigma(j)}-\mu_{\pi(j)}\right)}\prod_{N\geq j>k\geq 1}\left[\lambda_{\sigma(j),\sigma(k)}+ic^\prime \epsilon(z_j-z_k)\right]\left[\mu_{\sigma(j),\sigma(k)}-ic^\prime \epsilon(z_j-z_k)\right]\nonumber\\
\fl \times  \prod_{j=1}^N\left[\lambda_{\sigma(N+1),\sigma(j)} + ic^\prime \epsilon (x-z_j)\right]+
\sum_{\sigma \in S_{N+1}}\sum_{\pi \in S_{N}}(-1)^{\sigma +\pi}e^{-ix\lambda_{\sigma (N+1)}} 
\nonumber\\
\fl \int d^N z\, e^{-i\sum_{j=1}^N z_j\left(\lambda_{\sigma(j)}-\mu_{\pi(j)}\right)} \prod_{N\geq j>k\geq 1}\left[\lambda_{\sigma(j),\sigma(k)}+ic^\prime \epsilon(z_j-z_k)\right]\left[\mu_{\sigma(j),\sigma(k)}-ic^\prime \epsilon(z_j-z_k)\right]\nonumber\\
\fl \times  \left[\sum_{j=1}^N  \frac{d}{dx}\left\{ e^{-i (\pi\kappa /2)\epsilon(z_j-x)} \left[\lambda_{\sigma(N+1),\sigma(j)} + ic^\prime \epsilon (x-z_j)\right]\right\} \left( \prod_{k=1\atop k\neq j}^N e^{-i (\pi\kappa /2)\epsilon(z_k-x)}\right.\right.\nonumber\\ \left. \left. \times \left[\lambda_{\sigma(N+1),\sigma(k)} + ic^\prime \epsilon (x-z_k)\right]\right) \right]\,.
\label{eq:a3}
\eea
We now focus on the second term, and show
\bea
\sum_{r=1}^N\sum_{\sigma \in S_{N+1}}\sum_{\pi \in S_{N}}(-1)^{\sigma +\pi}e^{-ix\lambda_{\sigma (N+1)}}  \int d^N z\, e^{-i\sum_{j=1}^N z_j\left(\lambda_{\sigma(j)}-\mu_{\pi(j)}\right)}  \nonumber\\
\fl \frac{d}{dx}\left\{ e^{-i (\pi\kappa /2)\epsilon(z_r-x)} \left[\lambda_{\sigma(N+1),\sigma(r)} + ic^\prime \epsilon (x-z_r)\right]\right\} \left(\prod_{N\geq j>k\geq 1}\left[\lambda_{\sigma(j),\sigma(k)}+ic^\prime \epsilon(z_j-z_k)\right]\right.\nonumber\\
\fl \times \left.\left[\mu_{\sigma(j),\sigma(k)}-ic^\prime \epsilon(z_j-z_k)\right]\right)  \left[ \left( \prod_{k=1\atop k\neq r}^N e^{-i (\pi\kappa /2)\epsilon(z_k-x)}\left[\lambda_{\sigma(N+1),\sigma(k)} + ic^\prime \epsilon (x-z_k)\right]\right) \right]\nonumber\\
\fl =\sum_{r=1}^N\sum_{\sigma \in S_{N+1}}\sum_{\pi \in S_{N}}(-1)^{\sigma +\pi}e^{-ix\lambda_{\sigma (N+1)}}  \int d^N z\, e^{-i\sum_{j=1}^N z_j\left(\lambda_{\sigma(j)}-\mu_{\pi(j)}\right)}  \left[-i\left(\lambda_{\sigma(r)}-\mu_{\pi(r)}\right)\right] \nonumber\\
\times \left(\prod_{N\geq j>k\geq 1}\left[\lambda_{\sigma(j),\sigma(k)}+ic^\prime \epsilon(z_j-z_k)\right] \left[\mu_{\sigma(j),\sigma(k)}-ic^\prime \epsilon(z_j-z_k)\right]\right) \nonumber\\
 \times  \left ( \prod_{k=1}^N e^{-i (\pi\kappa /2)\epsilon(z_k-x)}\left[\lambda_{\sigma(N+1),\sigma(k)} + ic^\prime \epsilon (x-z_k)\right]\right) \,.
\label{eq:partial_2}
\eea
In order to do this, we proceed as follows. The r.h.s. is a sum of $N$ terms labeled by $r=1,\ldots N$.  Consider a generic term $r$, and perform integration over the variable $z_r$ in the multiple integral. The integral of the factors depending on $z_r$ yields
\bea
\fl \int_0^L d z_r  \prod_{k=1 \atop k\neq r}^{N}\left[\lambda_{\sigma(k),\sigma(r)}+ic^\prime \epsilon(z_k-z_r)\right]\left[\mu_{\sigma(k),\sigma(r)}-ic^\prime \epsilon(z_k-z_r)\right]
\nonumber\\ 
\times  e^{-i (\pi\kappa /2)\epsilon(z_r-x)}\left[\lambda_{\sigma(N+1),\sigma(r)} + ic^\prime \epsilon (x-z_r)\right] e^{-i z_r\left(\lambda_{\sigma(r)}-\mu_{\pi(r)}\right) }\nonumber\\
\fl   =\left\{\left( \prod_{k=1 \atop k\neq r}^{N}\left[\lambda_{\sigma(k),\sigma(r)}+ic^\prime \epsilon(z_k-z_r)\right]\left[\mu_{\sigma(k),\sigma(r)}-ic^\prime \epsilon(z_k-z_r)\right]\right. \right.\nonumber\\
\fl \left.\left.\left. \times e^{-i (\pi\kappa /2)\epsilon(z_r-x)}\left[\lambda_{\sigma(N+1),\sigma(r)} + ic^\prime \epsilon (x-z_r)\right] \right)\frac{e^{-i z_r(\lambda_{\sigma(r)}-\mu_{\pi(r)})}}{-i \left[\lambda_{\sigma(r)}-\mu_{\pi(r)}\right]}\right\}\right|^L_0\nonumber\\
\fl  -\int_0^L dz_r \frac{d}{dz_r}\left( 
\prod_{k=1 \atop k\neq r}^{N}\left[\lambda_{\sigma(k),\sigma(r)}+ic^\prime \epsilon(z_k-z_r)\right]\left[\mu_{\sigma(k),\sigma(r)}-ic^\prime \epsilon(z_k-z_r)\right]\right.\nonumber\\
\fl \left.\times  e^{-i (\pi\kappa /2)\epsilon(z_r-x)}\left[\lambda_{\sigma(N+1),\sigma(r)} + ic^\prime \epsilon (x-z_r)\right] 
\right)\frac{e^{-i z_r\left(\lambda_{\sigma(r)}-\mu_{\pi(r)}\right)}}{-i \left[\lambda_{\sigma(r)}-\mu_{\pi(r)}\right]}\,.
\label{eq:partial_1}
\eea
Crucially, it is easy to see that the first term is vanishing,  due to the Bethe equations~\eqref{eq:bethe_equations}. In order to evaluate the second term, we use the identity
\be
\frac{\mathrm{d} \epsilon\left(x-z_{n}\right)}{\mathrm{d} z_{n}}=-2 \delta\left(x-z_{n}\right)
\ee
so that the second integral in Eq.~\eqref{eq:partial_1} becomes
\bea
\fl  -\int_0^L dz_r \frac{d}{dz_r}\left( 
\prod_{k=1 \atop k\neq r}^{N}\left[\lambda_{\sigma(k),\sigma(r)}+ic^\prime \epsilon(z_k-z_r)\right]\left[\mu_{\sigma(k),\sigma(r)}-ic^\prime \epsilon(z_k-z_r)\right] \right.\nonumber\\
\fl \left.\times  e^{-i (\pi\kappa /2)\epsilon(z_r-x)} \left[\lambda_{\sigma(N+1),\sigma(r)} + ic^\prime \epsilon (x-z_r)\right] e^{-i z_r\left(\lambda_{\sigma(j)}-\mu_{\pi(j)}\right) }
\right)\frac{e^{-i z_r\left(\lambda_{\sigma(r)}-\mu_{\pi(r)}\right)}}{-i \left[\lambda_{\sigma(r)}-\mu_{\pi(r)}\right]}\nonumber\\
\fl = (\ast)_r + (\ast\ast)_r +(\ast\ast\ast)_r\,,
\eea
where
\bea
\fl (\ast)_r=-\int_0^L dz_r 
\prod_{k=1 \atop k\neq r}^{N}\left[\lambda_{\sigma(k),\sigma(r)}+ic^\prime \epsilon(z_k-z_r)\right]\left[\mu_{\sigma(k),\sigma(r)}-ic^\prime \epsilon(z_k-z_r)\right] \nonumber\\
\fl \times \frac{d}{dz_r}\left( e^{-i (\pi\kappa /2)\epsilon(z_r-x)} \left[\lambda_{\sigma(N+1),\sigma(r)} + ic^\prime \epsilon (x-z_r)\right] 
\right)\frac{e^{-i z_r\left(\lambda_{\sigma(r)}-\mu_{\pi(r)}\right)}}{-i \left[\lambda_{\sigma(r)}-\mu_{\pi(r)}\right]}\,,
\label{eq:star_def}
\eea
\bea
\fl (\ast\ast)_r=2i c^\prime \sum_{p=1\atop p\neq r}^N \left( e^{-i (\pi\kappa /2)\epsilon(z_p-x)} \left[\lambda_{\sigma(N+1),\sigma(r)} + ic^\prime \epsilon (x-z_p)\right] 
\right)\frac{e^{-i z_p\left(\lambda_{\sigma(r)}-\mu_{\pi(r)}\right)}}{-i \left[\lambda_{\sigma(r)}-\mu_{\pi(r)}\right]}\nonumber\\
\fl \times \left(\prod_{k=1 \atop k\neq r, p}^{N}\left[\lambda_{\sigma(k),\sigma(r)}+ic^\prime \epsilon(z_k-z_r)\right] \prod_{k=1 \atop k\neq r}^{N}\left[\mu_{\sigma(k),\sigma(r)}-ic^\prime \epsilon(z_k-z_r)\right]\right)\,,\\
\fl (\ast\ast\ast )_r = 2i c^\prime \sum_{p=1\atop p\neq r}^N \left( e^{-i (\pi\kappa /2)\epsilon(z_p-x)} \left[\lambda_{\sigma(N+1),\sigma(r)} + ic^\prime \epsilon (x-z_p)\right] 
\right)\frac{e^{-i z_p\left(\lambda_{\sigma(r)}-\mu_{\pi(r)}\right)}}{-i \left[\lambda_{\sigma(r)}-\mu_{\pi(r)}\right]}\nonumber\\
\fl \times \left(\prod_{k=1 \atop k\neq r}^{N}\left[\lambda_{\sigma(k),\sigma(r)}+ic^\prime \epsilon(z_k-z_r)\right] \prod_{k=1 \atop k\neq r, p}^{N}\left[\mu_{\sigma(k),\sigma(r)}-ic^\prime \epsilon(z_k-z_r)\right]\right)\,.
\eea
We can now plug these expressions into the r.h.s. of Eq.~\eqref{eq:partial_2}, and obtain
\bea
\fl \sum_{r=1}^N\sum_{\sigma \in S_{N+1}}\sum_{\pi \in S_{N}}(-1)^{\sigma +\pi}e^{-ix\lambda_{\sigma (N+1)}}  
\left(\prod_{s=1\atop s\neq r}^N\int d z_s\, e^{-i z_s\left(\lambda_{\sigma(s)}-\mu_{\pi(s)}\right)} \right) \left[-i\left(\lambda_{\sigma(r)}-\mu_{\pi(r)}\right)\right] \nonumber\\
\times \left(\prod_{N\geq j>k\geq 1\atop j,k\neq r}\left[\lambda_{\sigma(j),\sigma(k)}+ic^\prime \epsilon(z_j-z_k)\right] \left[\mu_{\sigma(j),\sigma(k)}-ic^\prime \epsilon(z_j-z_k)\right]\right)\nonumber\\
\times  \left[(\ast)_r+(\ast\ast)_r+(\ast\ast\ast)_r\right] \,.
\label{eq:almost_there}
\eea
Now, we claim  that the second and third terms, proportional to $(\ast\ast)$ and $(\ast\ast\ast)$ respectively, are vanishing. Let us consider for example the term proportional to $(\ast\ast)$. We have
\bea
\fl 2ic\sum_{r,p=1\atop r\neq p}^N \left(\prod_{s=1\atop s\neq r}^N\int d z_s\right) \sum_{\sigma \in S_{N+1}}\sum_{\pi \in S_{N}}(-1)^{\sigma +\pi}e^{-ix\lambda_{\sigma (N+1)}}   e^{-i\sum_{j\neq r,p}^N z_j\left(\lambda_{\sigma(j)}-\mu_{\pi(j)}\right)}  \nonumber\\
\fl \times  \left ( \prod_{k=1\atop k\neq r,p}^N e^{-i (\pi\kappa /2)\epsilon(z_k-x)}\left[\lambda_{\sigma(N+1),\sigma(k)} + ic^\prime \epsilon (x-z_k)\right]\right)\nonumber\\
\fl \times \left(\prod_{N\geq j>k\geq 1\atop k,r\neq r}\left[\lambda_{\sigma(j),\sigma(k)}+ic^\prime \epsilon(z_j-z_k)\right] \left[\mu_{\sigma(j),\sigma(k)}-ic^\prime \epsilon(z_j-z_k)\right]\right) \nonumber\\
\fl \times  e^{-i \pi\kappa \epsilon(z_p-x)}\left[\lambda_{\sigma(N+1),\sigma(p)} + ic^\prime \epsilon (x-z_p)\right] \left[\lambda_{\sigma(N+1),\sigma(r)} + ic^\prime \epsilon (x-z_p)\right]\nonumber\\
\times e^{-iz_p\left(\lambda_{\sigma(r)}+\lambda_{\sigma(p)}-\mu_{\sigma(r)}-\mu_{\sigma(p)}\right)}\,.
\eea
We claim that the terms in the sum over the permutations $\sigma\in S_{N+1}$ appear in pairs with opposite sign and equal absolute value, and thus cancel exactly. Indeed, for each permutation $\sigma$ consider the permutation $\tilde{\sigma}=\tau_{r,p}\circ \sigma$ where $\tau_{r,p}$ swaps $\lambda_p$ and $\lambda_r$ leaving stable $\lambda_j$ for $j\neq r,p$. Clearly, $(-1)^{\tilde{\sigma}}=-(-1)^\sigma$, so that the first line in the above sum gives opposite signs for the two permutations. On the other hand, lines $2$ through $5$ are exactly equal for the two terms, so that we have an overall minus sign, as anticipated.

With a similar reasoning (now involving the permutations $\pi$ instead of $\sigma$), we can prove that the term proportional to $(\ast\ast\ast)$ is also vanishing. Next, using 
\bea
 \frac{d}{dz_r}\left( e^{-i (\pi\kappa /2)\epsilon(z_r-x)} \left[\lambda_{\sigma(N+1),\sigma(r)} + ic^\prime \epsilon (x-z_r)\right]\right) \nonumber\\
= - \frac{d}{dx}\left( e^{-i (\pi\kappa /2)\epsilon(z_r-x)} \left[\lambda_{\sigma(N+1),\sigma(r)} + ic^\prime \epsilon (x-z_r)\right] 
\right)
\eea
and the definition in Eq.~\eqref{eq:star_def},  we see that Eq.~\eqref{eq:almost_there} implies Eq.~\eqref{eq:partial_2}.

Putting  Eqs.~\eqref{eq:a3} and \eqref{eq:partial_2} together, we finally obtain
\bea
\fl \frac{d}{dx}F_{N+1,N}(x)=\sum_{\sigma \in S_{N+1}}\sum_{\pi \in S_{N}}(-1)^{\sigma +\pi}e^{-ix\lambda_{\sigma (N+1)}}  \int d^N z\, e^{-i\sum_{j=1}^N z_j\left(\lambda_{\sigma(j)}-\mu_{\pi(j)}\right)}  \nonumber\\
\fl \times \left[-i\left(\sum_{r=1}^{N+1} \lambda_{\sigma(r)}-\sum_{r=1}^{N}\mu_{\pi(r)}\right)\right] 
 \left(\prod_{N\geq j>k\geq 1}\left[\lambda_{\sigma(j),\sigma(k)}+ic^\prime \epsilon(z_j-z_k)\right] \right.\nonumber\\
 \fl \times\left.\left[\mu_{\sigma(j),\sigma(k)}-ic^\prime \epsilon(z_j-z_k)\right]\right) \left ( \prod_{k=1}^N e^{-i (\pi\kappa /2)\epsilon(z_k-x)}\left[\lambda_{\sigma(N+1),\sigma(k)} + ic^\prime \epsilon (x-z_k)\right]\right)\nonumber\\
\fl =  \left[-i\left(\sum_{r=1}^{N+1} \lambda_{r}-\sum_{r=1}^{N}\mu_{r}\right)\right]  \sum_{\sigma \in S_{N+1}}\sum_{\pi \in S_{N}}(-1)^{\sigma +\pi}e^{-ix\lambda_{\sigma (N+1)}}  \int d^N z\, e^{-i\sum_{j=1}^N z_j\left(\lambda_{\sigma(j)}-\mu_{\pi(j)}\right)}  \nonumber\\
\fl \times \left(\prod_{N\geq j>k\geq 1}\left[\lambda_{\sigma(j),\sigma(k)}+ic^\prime \epsilon(z_j-z_k)\right] \left[\mu_{\sigma(j),\sigma(k)}-ic^\prime \epsilon(z_j-z_k)\right]\right) \nonumber\\
 \times  \left ( \prod_{k=1}^N e^{-i (\pi\kappa /2)\epsilon(z_k-x)}\left[\lambda_{\sigma(N+1),\sigma(k)} + ic^\prime \epsilon (x-z_k)\right]\right)
 \nonumber\\
 =  \left[-i\left(\sum_{r=1}^{N+1} \lambda_{r}-\sum_{r=1}^{N}\mu_{r}\right)\right] F_{N+1,N}(x)\,.
\eea

\section{Useful identities}
 \label{sec:identities}

In this appendix we discuss some identities involving sums of rational functions. The first useful identity is
\begin{equation}
	i\sum_{j=1}^{N+1}K^{+}(\lambda_s,\lambda_j)\frac{\prod_{m=1}^{N}(\mu_m-\lambda_j)}{\prod_{m\neq j}^{N+1}(\lambda_m-\lambda_j)}=-\left(V_{s}^+(\kappa=0)-V_{s}^-(\kappa=0)\right)\,,
	\label{eq:identity_1}
\end{equation}
where $V^{\pm}_s(\kappa)$ and $K^{+}(x,y)$ are defined in Eqs.~\eqref{eq:wvpm} and \eqref{eq:k_plus}. Eq.~\eqref{eq:identity_1} is obtained applying the residue theorem to the complex function
\begin{equation}
	g_s(z)=\frac{1}{(z-\lambda_s-ic^\prime)(z-\lambda_s+ic^\prime)}\frac{\prod_{m=1}^N(\mu_m-z)}{\prod_{m=1}^{N+1}(\lambda_m-z)}\,.
	\label{eq:complex_g1}
\end{equation}
Indeed the function $g_s(z)$ has first order poles for $z=\lambda_j$, $j=1,\ldots, \lambda_{N+1}$ and for $z=\lambda_{s}\pm ic$, while it is easy to see that it has vanishing residue at infinity. Using the fact that the sum of the residues has to be zero one immediately arrives at identity \eqref{eq:identity_1}.

The second useful identity is 
\begin{equation}
	i\sum_{j=1}^{N+1}K^{-}(\lambda_s,\lambda_j)\frac{\prod_{m=1}^{N}(\mu_m-\lambda_j)}{\prod_{m\neq j}^{N+1}(\lambda_m-\lambda_j)}=+\left(V_{s}^+(\kappa=0)+V_{s}^-(\kappa=0)\right)\,,
	\label{eq:identity_2}
\end{equation}
where $K^{-}(x,y)$ is defined in~\eqref{eq:k_minus}. Eq.~\eqref{eq:identity_2} is obtained applying again the residue theorem to the complex function
\begin{equation}
	g_s(z)=\frac{(z-\lambda_s)}{(z-\lambda_s-ic^\prime)(z-\lambda_s+ic^\prime)}\frac{\prod_{m=1}^N(\mu_m-z)}{\prod_{m=1}^{N+1}(\lambda_m-z)}\,.
	\label{eq:complex_g2}
\end{equation}

Finally, putting Eqs.~\eqref{eq:identity_1} and  \eqref{eq:identity_2} together, we arrive at the  third useful identity
\be
i\sum_{j=1}^{N+1}\mathcal{Q}_\kappa(\lambda_s,\lambda_j)\frac{\prod_{m=1}^{N}(\mu_m-\lambda_j)}{\prod_{m\neq j}^{N+1}(\lambda_m-\lambda_j)}=-\left(V_{s}^+(\kappa)-V_{s}^-(\kappa)\right)\,,
\label{eq:identity_3}
\ee
where $\mathcal{Q}_{\kappa}(x,y)$ is defined in \eqref{eq:q_function}.

\end{document}